\documentclass[12pt,notoc]{JHEP3}
\usepackage[english]{babel}
\usepackage{amsmath}
\usepackage{amsfonts}
\usepackage[dvips]{graphicx}
\usepackage{cite}
\newcommand{\starco}{\stackrel{*}{,}}
\newcommand{\inv}[1]{\frac{1}{#1}}

\def\r{\rho}
\def\s{\sigma}
\def\t{\tau}
\def\e{\epsilon}
\title{Gauge Independence of IR singularities in Non-Commutative QFT --- and Interpolating Gauges}
\author{M.~Attems\thanks{{\tt mattems@hep.itp.tuwien.ac.at}}~, D.N.~Blaschke\thanks{{\tt blaschke@hep.itp.tuwien.ac.at}}~, M.~Ortner\thanks{{\tt ortner@hep.itp.tuwien.ac.at}}~, M.~Schweda\thanks{{\tt mschweda@tph.tuwien.ac.at}}~, S.~Stricker\thanks{\tt stricker@hep.itp.tuwien.ac.at}~, M.~Weiretmayr\thanks{{\tt weiret@hep.itp.tuwien.ac.at}}\\
Institute for Theoretical Physics,\\Vienna University of Technology\\Wiedner Hauptstrasse 8-10, A-1040 Vienna, Austria\\\\
\footnotemark[4]~work supported by "Fonds zur F\"orderung der Wissenschaftlichen Forschung" (FWF) under contract P15463-N08\\
\footnotemark[1]~\footnotemark[2]~\footnotemark[3]~\footnotemark[5]~\footnotemark[6]~work supported by FWF under contract P15015-N08}
\abstract{IR divergences of a non-commutative U(1) Maxwell theory are discussed at the one-loop level using an interpolating gauge to show that quadratic IR divergences are independent not only from a covariant gauge fixing but also independent from an axial gauge fixing.}
\begin{document}
%%%%%%%%%%%%%%%%%%%%%%%%%%%%%%%%%%%%%%%%%
\section{Introduction}
%%%%%%%%%%%%%%%%%%%%%%%%%%%%%%%%%%%%%%%%%
In the perturbative realization of U(1) Maxwell gauge theory on non-commutative space-time the one-loop 1PI Green functions are known to have a non-analytic behaviour for small external momenta. The corresponding singularities are called non-commutative IR divergences and are present in the non-planar contributions of those Green functions whose planar counterparts are UV divergent by na\"ive power counting. This interplay between UV divergences and IR singularities is called the UV/IR mixing problem of non-commutative quantum field theories. It is interesting to stress that in non-commutative gauge field models, in using the so-called Seidberg-Witten map in order to construct $\theta$-expanded actions, the UV/IR mixing problem is absent~\cite{bichl1, bichl2, bichl3}.

A further nice feature is the fact that these IR singularities do not depend on the gauge parameter of covariant gauge-fixing of the corresponding gauge field models \cite{ruiz,armoni}. The aim of this paper is to also discuss the dependence on non-covariant non-standard gauge fixings --- like the axial gauges~\cite{boresch}. In order to study the gauge dependence of both cases simultaneously we use an interpolating gauge fixing~\cite{su-long, schweda, balasin} described in the next section.

We present the discussion of the vacuum polarization tensor at one-loop level and show that the quadratic IR divergences are in fact gauge independent in both cases: the covariant gauge and the axial gauge fixings~\cite{Kummer}.
%%%%%%%%%%%%%%%%%%%%%%%%%%%%%%%%%%%%%%%%%
\section{Interpolating gauge fixing and non-commu\-tative U(1) \\Max\-well theory at the classical level}\label{sect2}
%%%%%%%%%%%%%%%%%%%%%%%%%%%%%%%%%%%%%%%%%
For a wide class of linear gauges --- i.e. the standard one and the non-standard axial gauge --- the gauge independence of these IR singularities is a signal for a new kind of physics. In general, physical quantities are gauge invariant --- meaning BRS invariance here --- with no dependence on the ghost fields and the gauges chosen to quantize the U(1) gauge field model. However, one has to stress that the one-loop correction to the vacuum polarization is a 1PI vertex which is only the building element for the construction of connected physical contributions.

The discussion of non-commutative quantum field theories is based on Filk's proposal to replace the products of fields in any action by the so-called Weyl-Moyal products~\cite{filk}. In the case of a U(1) Maxwell theory one has the following replacement for the product of photon fields $A_\mu$:
\begin{align}
A_\mu(x) A_\nu (x) \rightarrow A_\mu(x) \star A_\nu(x),
\end{align}
with the definition
\begin{align}\label{star_prod_gauge_fields}
A_\mu(x) \star A_\nu ( x ) = 
e^{\frac{i}{2}\theta^{\mu\nu}\partial^x_\mu\partial^y_\nu} A_\mu(x) A_\nu(y)\Big
|_{x = y}\,,
\end{align}
where $\theta^{\mu\nu}$ is the constant antisymmetric deformation parameter with $-2$ in mass dimension (for natural units $\hbar = c = 1$). The equation (\ref{star_prod_gauge_fields}) implies that the star commutator of two commuting space-time coordinates becomes
\begin{align}
\left[ x^\mu \starco x^\nu \right] = x^\mu \star x^\nu - x^\nu \star x^\mu 
= i \theta^{\mu\nu},
\end{align}
Due to the fact that the product (\ref{star_prod_gauge_fields}) is non-commutative, the extended Maxwell field strength is given by~\cite{ruiz, hayakawa}
\begin{align}
F_{\mu\nu} = \partial_\mu A_\nu - \partial_\nu A_\mu - ig [ A_\mu \starco A_\nu].
\label{electro_magnetic_fieldtensor}
\end{align}
With (\ref{electro_magnetic_fieldtensor}) one can define the following gauge invariant action at the classical level:
\begin{align}
\Gamma_{\text{INV}} = - \inv 4 \int d^4x
F_{\mu\nu} \star F^{\mu\nu},
\end{align}
with a gauge transformation of the form
\begin{align}
\delta_\lambda A_\mu = \partial_\mu \Lambda - ig[A_\mu\starco\Lambda],
\end{align}
where $\Lambda$ is infinitesimal. Quantization, especially the calculation of the gauge field propagator, enforces the breaking of local gauge invariance. This is done in fixing the gauge by using the constraint
\begin{align}
N_\mu A^\mu = 0,&&\text{with}&& N_\mu = \partial_\mu - \xi \frac{(n \partial)}{n^2} n_\mu,
\label{interpolating_gauge}
\end{align}
which was originally proposed in~\cite{su-long} and used in~\cite{schweda, balasin}. The constant vector $n^\mu$ and the real variable $\xi$ are gauge parameters --- $\xi$ taking values between $(-\infty, +1)$. This allows to interpolate between a linear class of gauges: the covariant one ($\xi=0$) and the axial gauge ($\xi\to-\infty$).

In order to quantize non-commutative U(1) Maxwell theory consistently one has to use the BRS procedure entailing the introduction of the Faddeev-Popov ($\Phi \Pi$) ghost fields $\bar c$ (antighost field) and $c$ (ghost field):
\begin{align}
\Gamma^{(0)'} =  \Gamma_{\text{INV}} + \Gamma_ {\Phi \Pi}^{'} = \int d^4x
\left[ - \inv 4 F_{\mu\nu} \star F^{\mu\nu} +
B \star N^\mu A_\mu - \bar c\star N^\mu D_\mu c \right],
\label{action simple}
\end{align}
where $B$ is the multiplier field to implement the gauge constraint (\ref{interpolating_gauge}). Additionally, in order to be more general, we can also introduce a further gauge parameter $\alpha$ changing (\ref{action simple}) into
\begin{align}
\Gamma^{(0)} =  \Gamma_{\text{INV}} + \Gamma_ {\Phi \Pi} = \int d^4x
\left[ - \inv 4 F_{\mu\nu} \star F^{\mu\nu} + \frac{\alpha}{2} B \star B +
B \star N^\mu A_\mu - \bar c\star N^\mu D_\mu c \right].
\label{action}
\end{align}
The actions (\ref{action simple}),(\ref{action}) are invariant with respect to the BRS symmetry defined by \cite{Piguet}
\begin{align}
& sA_\mu = D_\mu c= \partial_\mu c - ig [A_\mu \starco c], && sc=igc \star c, \nonumber\\
& s \bar c = B, && sB = 0, \nonumber \\
&s^2\phi=0, \hspace{1cm} \text{for}\ \phi=\{A^{\mu},B,c,\bar{c}\}. 
\label{brs}
\end{align}
The transformations (\ref{brs}) are nilpotent, non-linear and supersymmetric. For describing the symmetry content encoded by equations (\ref{brs}) one has to add a term of the following form to equation (\ref{action}):
\begin{align}
\Gamma_{\text{ex}} = \int d^4x 
\left[ \rho^\mu \star s A_\mu + \sigma \star sc \right],
\end{align}
where $\rho^\mu$ and $\sigma$ are unquantized external BRS invariant sources for the non-linear contributions of the BRS-transformations. The symmetry content of
\begin{align}
\Gamma^{(0)} = \Gamma_{\text{INV}} + \Gamma_{\Phi \Pi} + \Gamma_{\text{ex}},
\end{align}
is now described by the non-linear Slavnov identity
\begin{align}
\mathcal{S}\left(\Gamma^{(0)}\right) =
\int d^4x \left(\frac{\delta \Gamma^{(0)}}{\delta \rho^\mu} \star
\frac{\delta \Gamma^{(0)}}{\delta A_\mu} +
\frac{\delta \Gamma^{(0)}}{\delta \sigma} \star
\frac{\delta \Gamma^{(0)}}{\delta c} +
B \star \frac{\delta \Gamma^{(0)}}{\delta \bar c}\right) = 0.
\label{slavnov_identity}
\end{align}
The use of the star product (\ref{star_prod_gauge_fields}) in the bilinear action has no effect. Thus, the free field theory remains unchanged and therefore the propagators of the U(1) Maxwell theory are not touched by non-commutativity.

In momentum representation the gauge field propagator becomes
\begin{align}
i\Delta^{AA}_{\mu\nu} (k) = -\frac{i}{k^2} \left[ g_{\mu\nu}
- a k_\mu k_\nu  + b (n_\mu k_\nu + n_\nu k_\mu)
\right],
\label{gauge_field_propagator}
\end{align}
with 
\begin{align}\label{defa}
a  = \frac{(1-\alpha)k^2-\zeta^2n^2(nk)^2}{\left[k^2-\zeta(nk)^2\right]^2},
\end{align}
and
\begin{align}\label{defb}
b = \frac{\zeta(nk)}{k^2-\zeta(nk)^2},
\end{align}
where $\zeta=\frac{\xi}{n^2}$.
In the limit $\zeta \rightarrow 0$ $(\xi \rightarrow 0)$ one recovers the usual gauge field propagator for a covariant gauge fixing:
\begin{align}
i\Delta^{cov}_{\mu \nu} = -\frac{i}{k^2} \left[g_{\mu \nu} - (1 - \alpha) 
\frac{k_\mu k_\nu}{k^2} \right].
\end{align}
In the limit $\zeta \rightarrow - \infty$ ($\xi \rightarrow - \infty$) and $n^2 \ne 0$ one has the corresponding gauge field propagator in the axial gauge:
\begin{align}
i\Delta^{ax}_{\mu \nu} =  -\frac{i}{k^2} \left[g_{\mu \nu} -
\frac{n_\mu k_\nu + n_\nu k_\mu}{(nk)} +
n^2 \frac{k_\mu k_\nu}{(nk)^2} \right].
\end{align}
The remaining ghost-antighost propagator is given by
\begin{align}\label{ghostprop}
i\Delta_{c \bar c} =  \frac{i}{k^2 - \zeta(nk)^2},
\end{align}
and the mixed propagator between the gauge field $A_\mu$ and the multiplier  field $B$ becomes
\begin{align}\label{bprop}
i\Delta_{\mu B} = - \frac{k_\mu}{k^2 - \zeta (nk)^2}.
\end{align}
One observes that also the propagators (\ref{ghostprop}) and (\ref{bprop}) depend on the gauge parameter $\zeta$. Additionally, one has to mention that the vertex for the interaction of the gauge field and the ghosts is gauge dependent as well:
\begin{align}\label{ghostvertex}
V_{\mu c \bar c}(q_1, q_2, k)&=2g\left(q_{2\mu}-\zeta(nq_2)n_\mu\right)\sin\left(\frac{
q_1\tilde{q}_2}{2}\right),
\end{align}
where $\tilde q_2^\mu$ is defined by $\tilde q_2^\mu = \theta^{\mu \nu} q_{2\nu}$. $k_\mu$ denotes the gauge field momentum and $q_{i\mu}$ ($i = 1,2$) are the momenta of the ghost fields.

The other couplings describing the self-interactions of the bosons (stemming from the invariant part of the action) are gauge independent and are well-known in the literature~\cite{ruiz, armoni, hayakawa}. The three-photon vertex is given by
\begin{align}\label{3Avertex}
V^{3A}_{\rho\sigma\tau}(k_1, k_2, k_3)=&-2g \Big[(k_3-k_2)_\rho g_{\sigma\tau}+(
k_1-k_3)_\sigma g_{\rho\tau}+ \nonumber \\
&\quad\quad + (k_2-k_1)_\tau g_{\rho\sigma}\Big]\sin\left(\frac
{k_1\tilde{k}_2}{2}\right),
\end{align}
and the four-boson vertex is
\begin{align}\label{4Avertex}
V^{4A}_{\rho\sigma\tau\epsilon}(k_1, k_2, k_3, k_4)&=-4ig^2\left[(g_{\rho\tau}g_
{\sigma\epsilon}-g_{\rho\epsilon}g_{\sigma\tau})\sin\left(\frac{k_1\tilde{k}_2}{
2}\right)\sin\left(\frac{k_3\tilde{k}_4}{2}\right)\right.\nonumber\\
&\quad\qquad\left.+(g_{\rho\sigma}g_{\tau\epsilon}-g_{\rho\epsilon}g_{\sigma\tau
})\sin\left(\frac{k_1\tilde{k}_3}{2}\right)\sin\left(\frac{k_2\tilde{k}_4}{2} \right)\right.\nonumber\\
&\quad\qquad\left.+(g_{\rho\sigma}g_{\tau\epsilon}-g_{\rho\tau}g_{\sigma\epsilon
})\sin\left(\frac{k_2\tilde{k}_3}{2}\right)\sin\left(\frac{k_1\tilde{k}_4}{2}\right)\right].
\end{align}
One observes that the Feynman rules for the vertices contain phases. Due to this fact the behaviour for high internal momenta of the corresponding Feynman integrals in momentum representation is modified in a new fashion: For high internal momenta the phases act as a regularization induced by the oscillating phase factors. This implies that non-planar 1PI graphs, which are \`a priori UV divergent by na\"ive power counting, become finite but develop a new singularity for small external momenta. This interplay between the expected UV divergences --- which are not present --- and the existence of the real IR singularity represents the so-called UV/IR mixing problem~\cite{Minwalla, Susskind}. This UV/IR mixing problem is a one-loop effect and leads to inconsistencies in higher loop order corrections and therefore presents a severe obstacle for the renormalization program of any non-commutative quantum field theory.
%%%%%%%%%%%%%%%%%%%%%%%%%%%%%%%%%%%%%%%%%
\section{Gauge-fixing independence of IR divergences in the realm of the interpolating gauge fixing at one-loop level}
%%%%%%%%%%%%%%%%%%%%%%%%%%%%%%%%%%%%%%%%%
In this section we want to discuss the one-loop corrections to the photon vacuum polarization $\Pi_{\mu\nu}(p)$ in the framework of the interpolating gauge mentioned above. The aim of our investigation is to show that the corresponding IR singularities (induced by the deformation) are gauge fixing independent. From the discussion of the non-planar tadpole graph at one-loop level (see Fig.~\ref{fig:scalar-tadpole}) for a scalar field model
one has the following contribution
\begin{align}
\Gamma^{(1)} (p) \cong \int \frac{d^4k}{(2 \pi)^4} \inv{k^2} e^{ik \tilde p}
\cong \inv{\tilde p^2},
\label{tadpole_one}
\end{align}
where we have neglected the mass. By na\"ive power counting the 1PI graph (\ref{tadpole_one}) has a quadratic divergence for high internal momentum $k$. However, the result shows a finite expression --- but which is singular for small external momentum. Without considering the usual technical details for the calculation of Feynman graphs the result (\ref{tadpole_one}) can be easily understood by dimensional analysis.
\FIGURE[h]{\parbox[h]{15cm}{\centering\includegraphics[scale=1]{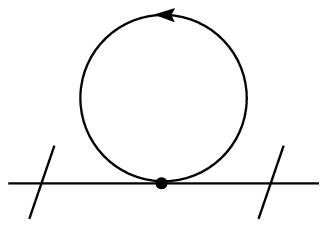}\caption{amputated scalar tadpole graph}\label{fig:scalar-tadpole}}}

The aim of this section is to investigate the gauge independence of IR singularities emerging from the one-loop corrections to the vacuum polarization. For this reason one has to consider the following three amputated one-loop graphs presented in Fig.~\ref{fig:photonself}.
\FIGURE[ht]{\parbox[h]{15cm}{\centering\includegraphics[scale=1]{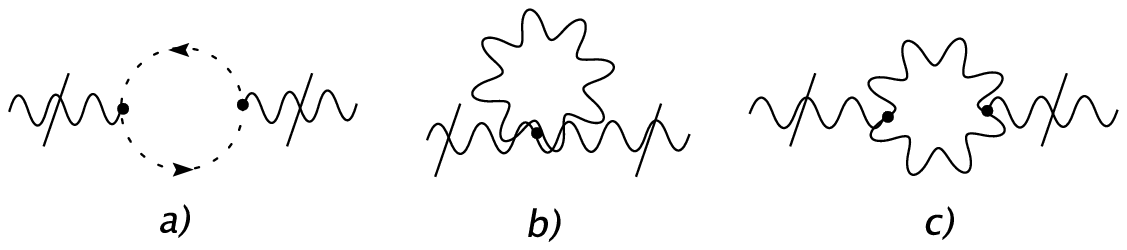}\caption{gauge boson self-energy --- amputated graphs}\label{fig:photonself}}}

Corresponding to the Feynman rules given in Section~\ref{sect2} the vacuum polarization tensor $\Pi_{\mu\nu}(p)$ in the one-loop approximation is a Feynman integral of the following form:
\begin{align}
i\Pi_{\mu\nu} (p) = \int d^4k I_{\mu\nu} (k, p) \sin^2 \left(\frac{k\tilde{p}}{2}\right).
\end{align}
Details are given in the Appendix. Additionally, also the transversality condition
\begin{align}
p^\mu \Pi_{\mu\nu}(p) = 0,
\end{align}
follows from the Slavnov identity (\ref{slavnov_identity}).

In order to isolate the expected IR singularities of the non-planar sector, one proceeds as in the standard renormalization program for planar graphs (i.e. graphs without phases) in considering the following expansion
\begin{align}
i\Pi_{\mu\nu} (p) =& \int d^4k I_{\mu \nu} (k, 0) \sin^2 \left(\frac{k\tilde{p}}{2}\right) + 
p^\rho \int d^4k \frac{\partial}{\partial p^\rho} I_{\mu \nu} (k , 0)
 \sin^2 \left(\frac{k\tilde{p}}{2}\right) + \nonumber\\
& + \inv 2 p^\sigma p^\rho \int d^4k 
\frac{\partial^2}{\partial p^\sigma \partial p^\rho} I_{\mu \nu} (k , 0)
 \sin^2 \left(\frac{k\tilde{p}}{2}\right)+\ldots 
 \label{expansion}
\end{align}
Due to the fact that the na\"ive degree of divergence $D = 2$ ($D= 4 -E =2$, $E$ external bosons) for high internal $k$ one is inclined to believe that the first term of (\ref{expansion}) is a candidate for a quadratic non-commutative IR singularity. The second may be linearly divergent. However, for dimensional reasons no linear IR divergences occur. The third term in (\ref{expansion}) may contain logarithmic divergences.

Calculation of the first term of (\ref{expansion}) leads to
\begin{align}
 \int d^4k I^{\mu \nu} (k, 0) &\sin^2 \left(\frac{k\tilde{p}}{2}\right) =4g^2\int\frac{d^4k}{(2\pi)^4}\sin^2\left(\frac{k\tilde{p}}{2}\right)\inv{k^2}\bigg\{-2g^{\mu\nu}-\nonumber\\
&\quad-b\left(n^\mu k^\nu+k^\mu n^\nu\right)\left[1+(nk)b-\frac{k^2}{k^2-\zeta(nk)^2}\right]+\nonumber\\
&\quad+\frac{k^\mu k^\nu}{k^2}\left[5+2(nk)b+(nk)^2b^2-\frac{k^4}{\left[k^2-\zeta(nk)^2\right]^2}\right]\bigg\},
\end{align}
(cf. Appendix). This expression is obviously independent of $a$, which was defined in (\ref{defa}), and hence independent of $\alpha$. With the definition (\ref{defb}) one finally gets
\begin{align}
 \int d^4k I_{\mu \nu} (k, 0) \sin^2 \left(\frac{k\tilde{p}}{2}\right) =
 4g^2 \int \frac{d^4k}{(2\pi)^4} \left[ 4 \frac{k_\mu k_\nu}{k^4} 
 - 2 \frac{g_{\mu \nu}}{k^2} \right] \sin^2 \left( \frac{k \tilde p}{2} \right).
\end{align}
One observes that the gauge dependent tensor structure based on the existence of the gauge directions $n_\mu$ and the dependence on the gauge parameters $\alpha$ and $\zeta$ (or $\xi$ with $\zeta=\xi/n^2$) cancel completely. Performing the integration we reproduce the known result~\cite{ruiz, hayakawa}
\begin{align}\label{result}
\int d^4k I_{\mu \nu} (k, 0) \sin^2 \left(\frac{k\tilde{p}}{2}\right) = i\frac{2g^2}{\pi^2}\frac{\tilde{p}_\mu\tilde{p}_\nu}{\tilde{p}^4}.
\end{align}
The final result of the non-planar contributions to the vacuum polarization at the one-loop level is finite but shows the expected quadratic IR singularity for small external momentum. The pole term in equation (\ref{result}) is the manifestation of the so-called UV/IR mixing that is a typical new feature of non-commutative quantum field theories. The origin of these singularities is the UV regime which seems to influence the IR behaviour of the field model. These pole terms create serious problems in the renormalization procedure, since when the non-planar singular IR contributions are inserted into higher loop diagrams they create new divergences. One possible way to bypass these difficulties is to use the Slavnov trick~\cite{slavnov1, slavnov2}.
%%%%%%%%%%%%%%%%%%%%%%%%%%%%%%%%%%%%%%%%%%%%%%%%%%%%%%%%
\begin{appendix}
\section*{Appendix}
\setcounter{equation}{0}
\renewcommand{\theequation}{\roman{equation}}
%%%%%%%%%%%%%%%%%%%%%%%%%%%%%%%%%%%%%%%%%%%%%%%%%%%%%%%%
The gauge boson self-energy at the one-loop level consists of three graphs depicted in Figure~\ref{fig:photonself}: The ghost loop $\Pi^{\mu\nu}_a(p)$ (Fig.~\ref{fig:photonself}a), the tadpole graph $\Pi^{\mu\nu}_b(p)$ (Fig.~\ref{fig:photonself}b) and the boson loop $\Pi^{\mu\nu}_c(p)$ (Fig.~\ref{fig:photonself}c). The first term in the expansion (\ref{expansion}) is then given by
\begin{align}\label{expansion-appendix}
\int d^4k I^{\mu \nu} (k, 0)& \sin^2 \left(\frac{k\tilde{p}}{2}\right) = \int d^4k I^{\mu \nu}_a (k, 0) \sin^2 \left(\frac{k\tilde{p}}{2}\right) +\nonumber\\
&\quad+\int d^4k I^{\mu \nu}_b (k, 0) \sin^2 \left(\frac{k\tilde{p}}{2}\right) +\int d^4k I^{\mu \nu}_c (k, 0) \sin^2 \left(\frac{k\tilde{p}}{2}\right) \equiv\nonumber\\
&\quad\equiv i\Pi^{\mu\nu}_{a,IR}(p)+i\Pi^{\mu\nu}_{b,IR}(p)+i\Pi^{\mu\nu}_{c,IR}(p).
\end{align}
According to the Feynman rules (\ref{ghostprop}),(\ref{ghostvertex}) the ghost-loop contribution is
\begin{align}
i\Pi^{\mu\nu}_{a,IR}&(p)=4g^2\int\frac{d^4k}{(2\pi)^4}\sin^2\left(\frac{k\tilde{p}}{2}\right)\frac{-\left[k^\mu-\zeta(nk)n^\mu\right]\left[k^\nu-\zeta(nk)n^\nu\right]}{\left[k^2-\zeta(nk)^2\right]^2}=\nonumber\\
&=4g^2\int\frac{d^4k}{(2\pi)^4}\sin^2\left(\frac{k\tilde{p}}{2}\right)\left\{\frac{-k^\mu k^\nu}{\left[k^2-\zeta(nk)^2\right]^2}+b\frac{\left(n^\mu k^\nu+k^\mu n^\nu\right)}{k^2-\zeta(nk)^2}-b^2n^\mu n^\nu\right\},
\end{align}
where the abbreviations defined in (\ref{defa}),(\ref{defb}) were used. The tadpole contribution to (\ref{expansion-appendix}) (see (\ref{gauge_field_propagator}),(\ref{4Avertex})) is
\begin{align}
i\Pi^{\mu\nu}_{b,IR}(p)&=2g^2\int\frac{d^4k}{(2\pi)^4}\sin^2\left(\frac{k\tilde{p}}{2}\right)\left(g^{\mu\tau}g^{\s\nu}+g^{\mu\s}g^{\tau\nu}-2g^{\mu\nu}g^{\s\tau}\right)\times\nonumber\\
&\quad\qquad\qquad\times\inv{k^2}\left[g_{\t\s}-ak_\t k_\s+b(n_\t k_\s+k_\t n_\s)\right]\nonumber\\
&=4g^2\int\frac{d^4k}{(2\pi)^4}\sin^2\left(\frac{k\tilde{p}}{2}\right)\inv{k^2}\big\{g^{\mu\nu}\left[k^2a-3-2(nk)b\right]+\nonumber\\
&\quad\qquad\qquad+b\left(n^\mu k^\nu+k^\mu n^\nu\right)-ak^\mu k^\nu\big\}.
\end{align}
Finally, the contribution of the photon loop according to Feynman rules (\ref{gauge_field_propagator}),(\ref{3Avertex}) is
\begin{align}
i\Pi^{\mu\nu}_{c,IR}(p)=2g^2\int&\frac{d^4k}{(2\pi)^4}\sin^2\left(\frac{k\tilde{p}}{2}\right)\inv{k^4}\left[-k^\e g^{\mu\s}+2k^\mu g^{\e\s}-k^\s g^{\e\mu}\right]\times\nonumber\\
\times&\left[g_{\t\e}-ak_\t k_\e+b(n_\t k_\e+k_\t n_\e)\right]\left[-k^\r g^{\nu\t}+2k^\nu g^{\r\t}-k^\t g^{\r\nu}\right]\times\nonumber\\
\times&\left[g_{\s\r}-ak_\s k_\r+b(n_\s k_\r+k_\s n_\r)\right].
\end{align}
Noticing that
\begin{align}
&\left[-k^\e g^{\mu\s}+2k^\mu g^{\e\s}-k^\s g^{\e\mu}\right]\left[g_{\t\e}-ak_\t k_\e+b(n_\t k_\e+k_\t n_\e)\right]=\nonumber\\
&=\big[-k_\t g^{\mu\s}+2k^\mu g_\t^{\ \s}-k^\s g_\t^{\ \mu}+ak_\t(k^2g^{\mu\s}-k^\mu k^\s)+bn_\t(k^\mu k^\s-k^2g^{\mu\s})+\nonumber\\
&\quad+bk_\t(-nkg^{\mu\s}+2k^\mu n^\s-n^\mu k^\s)\big]\nonumber\\
&=\big[fk_\t g^{\mu\s}-k_\t k^\s(ak^\mu+bn^\mu)+2k^\mu g_\t^{\ \s}-k^\s g_\t^{\ \mu}+\nonumber\\
&\quad+bn_\t(k^\mu k^\s-k^2g^{\mu\s})+2bk^\mu k_\t n^\s\big],
\end{align}
with the abbreviation
\begin{align}\label{deff}
f=k^2a-1-(nk)b,
\end{align}
we get
\begin{align}
i\Pi^{\mu\nu}_{c,IR}(p)&=2g^2\int\frac{d^4k}{(2\pi)^4}\sin^2\left(\frac{k\tilde{p}}{2}\right)\inv{k^4}\big[fk_\t g^{\mu\s}-k_\t k^\s(ak^\mu+bn^\mu)+2k^\mu g_\t^{\ \s}-\nonumber\\
&-k^\s g_\t^{\ \mu}+bn_\t(k^\mu k^\s-k^2g^{\mu\s})+2bk^\mu k_\t n^\s\big]\big[fk_\s g^{\nu\t}+2k^\nu g_\s^{\ \t}-\nonumber\\
&-k_\s k^\t(ak^\nu+bn^\nu)-k^\t g_\s^{\ \nu}+bn_\s(k^\nu k^\t-k^2g^{\nu\t})+2bk^\nu k_\s n^\t\big],
\end{align}
leading to
\begin{align}
i\Pi^{\mu\nu}_{c,IR}(p)=2g^2&\int\frac{d^4k}{(2\pi)^4}\sin^2\left(\frac{k\tilde{p}}{2}\right)\inv{k^4}\big\{2k^4b^2n^\mu n^\nu+2k^2g^{\mu\nu}\left[(nk)b-f\right]+\nonumber\\
&\quad+k^2b\left(n^\mu k^\nu+k^\mu n^\nu\right)\left[k^2a-f-5-3(nk)b\right]+\nonumber\\
&\quad+k^\mu k^\nu\big[f^2-2k^2af+4f+4(nk)bf+k^4a^2-2k^2a-\nonumber\\
&\quad\qquad\quad-4k^2(nk)ab+4d-3+10(nk)b+5(nk)^2b^2\big]\big\},
\end{align}
where $d=tr(g^{\mu\nu})=4$. Using (\ref{deff}) this expression becomes
\begin{align}
i\Pi^{\mu\nu}_{c,IR}(p)&=4g^2\int\frac{d^4k}{(2\pi)^4}\sin^2\left(\frac{k\tilde{p}}{2}\right)\inv{k^2}\bigg\{k^2b^2n^\mu n^\nu-b\left(n^\mu k^\nu+k^\mu n^\nu\right)\left[2+(nk)b\right]-\nonumber\\
&\quad-g^{\mu\nu}\left[k^2a-1-2(nk)b\right]+\frac{k^\mu k^\nu}{k^2}\left[k^2a+5+2(nk)b+(nk)^2b^2\right]\bigg\}.
\end{align}
The sum of these three graphs (\ref{expansion-appendix}) is then
\begin{align}\label{result-appendix}
\int d^4k I^{\mu \nu} (k, 0) \sin^2 \left(\frac{k\tilde{p}}{2}\right) &=4g^2\int\frac{d^4k}{(2\pi)^4}\sin^2\left(\frac{k\tilde{p}}{2}\right)\inv{k^2}\bigg\{-2g^{\mu\nu}-\nonumber\\
&\quad-b\left(n^\mu k^\nu+k^\mu n^\nu\right)\left[1+(nk)b-\frac{k^2}{k^2-\zeta(nk)^2}\right]+\nonumber\\
&\quad+\frac{k^\mu k^\nu}{k^2}\left[5+2(nk)b+(nk)^2b^2-\frac{k^4}{\left[k^2-\zeta(nk)^2\right]^2}\right]\bigg\}\nonumber\\
&=4g^2\int\frac{d^4k}{(2\pi)^4}\left\{4\frac{k^\mu k^\nu}{k^4}-2\frac{g^{\mu\nu}}{k^2}\right\}\sin^2\left(\frac{k\tilde{p}}{2}\right).
\end{align}
\end{appendix}
%%%%%%%%%%%%%%%%%%%%%%%%%%%%%%%%%%%%%%%%%%%%%%%%%%%%%%%

\end{document}